\documentclass[twocolumn,secnumarabic,amssymb, nobibnotes,aps,prl,superscriptaddress]{revtex4-2}
\usepackage{graphicx}
\usepackage{textcomp}
\usepackage{amsmath}
\usepackage{commath}
\usepackage{color}
\usepackage{appendix}

\newcommand{\sg}[1]{}
\renewcommand{\sg}[1]{{\color{red}{#1}}} % Comment out to leave notes out (final version)
\begin{document}

\title{Robustness of stress focusing in soft lattices under topology-switching deformation}
\author{Caleb Widstrand}
\affiliation{Department of Civil, Environmental, and Geo- Engineering, University of Minnesota, Minneapolis, MN 55455, USA}
\author{Xiaoming Mao}
\affiliation{Department of Physics, University of Michigan, Ann Arbor, MI 48109, USA}
%\author{Joseph Labuz}
%\affiliation{Department of Civil, Environmental, and Geo- Engineering, University of Minnesota, Minneapolis, MN 55455, USA}
\author{Stefano Gonella}
\email{sgonella@umn.edu}
\affiliation{Department of Civil, Environmental, and Geo- Engineering, University of Minnesota, Minneapolis, MN 55455, USA}

\begin{abstract}
    \noindent 
    Recent developments in topological mechanics have demonstrated the ability of Maxwell lattices to effectively focus stress along domain walls between differently polarized domains. The focusing ability can be exploited to protect the lattice bulk from accidental stress concentration - and eventually onset and propagation of fracture - at structural hot spots such as defects and cracks. A recent study has revisited the problem for structural lattices featuring non-ideal hinges, showing that the focusing remains robust, albeit diluted in strength. Realizing that the problem of domain wall localization has been traditionally framed in the context of linear elasticity, in this work we extend the study to the realm of soft structures undergoing nonlinear finite deformation. Through experiments performed on silicone hyperelastic prototypes, we assess and quantify the robustness of the phenomenon against the macroscopic shape changes induced by large deformation, with special attention to deformation levels that alter the topology of the bulk, lifting the topological protection. Furthermore, we identify a simple geometric indicator for this transition.
    \vspace{0.4cm}
\end{abstract}	
\maketitle

%\section{Introduction}

Mechanical metamaterials are architected solids whose microstructure is engineered to yield unconventional mechanical properties~\cite{CuiMeta_2010}. Recently, the intersection of topological mechanics and metamaterial design has led to the discovery of new classes of mechanical metamaterials whose response is attributable to the topology of their bulk structure. The topological characterization can be carried out in physical space, involving the shapes of the geometrical layout, or in the so-called \textit{k-space}, in terms of descriptors buried in the system's phonon band structure. An example of systems whose properties are dictated by their k-space topology is topological Maxwell lattices. In the ideal configuration, where the sites act as perfect hinges, Maxwell lattices possess an equal number of constraints and degrees of freedom in the bulk~\cite{Maxwell_1864}. Typical examples in two dimensions are the kagome and the square lattices. \textit{Topological} Maxwell lattices are a subclass of Maxwell configurations featuring topological \textit{polarization}, which grants them the ability to localize deformation (\textit{floppy modes}) on selected edges or interfaces of finite domains. This results in an asymmetric mechanical response~\cite{KL_2013,Rocklin_2017,Rocklin_2017_NJP,Bilal_2017,ZhouZhangMao_Fibers_2018,Mao_Lubensky_2018,Baardink_PNAS_2018}, whereby the excess of ``floppiness" on one edge is matched by a surplus of rigidity on the opposite edge~\cite{KL_2013,Bilal_2017,Guo_2019,Pishvar_Harne_2020}. 
The polarization is captured by a polarization vector, whose orientation marks the direction along which floppy modes exponentially localize~\cite{KL_2013}.

\begin{figure*}[t]
	\centering
	\includegraphics[width=1\textwidth]{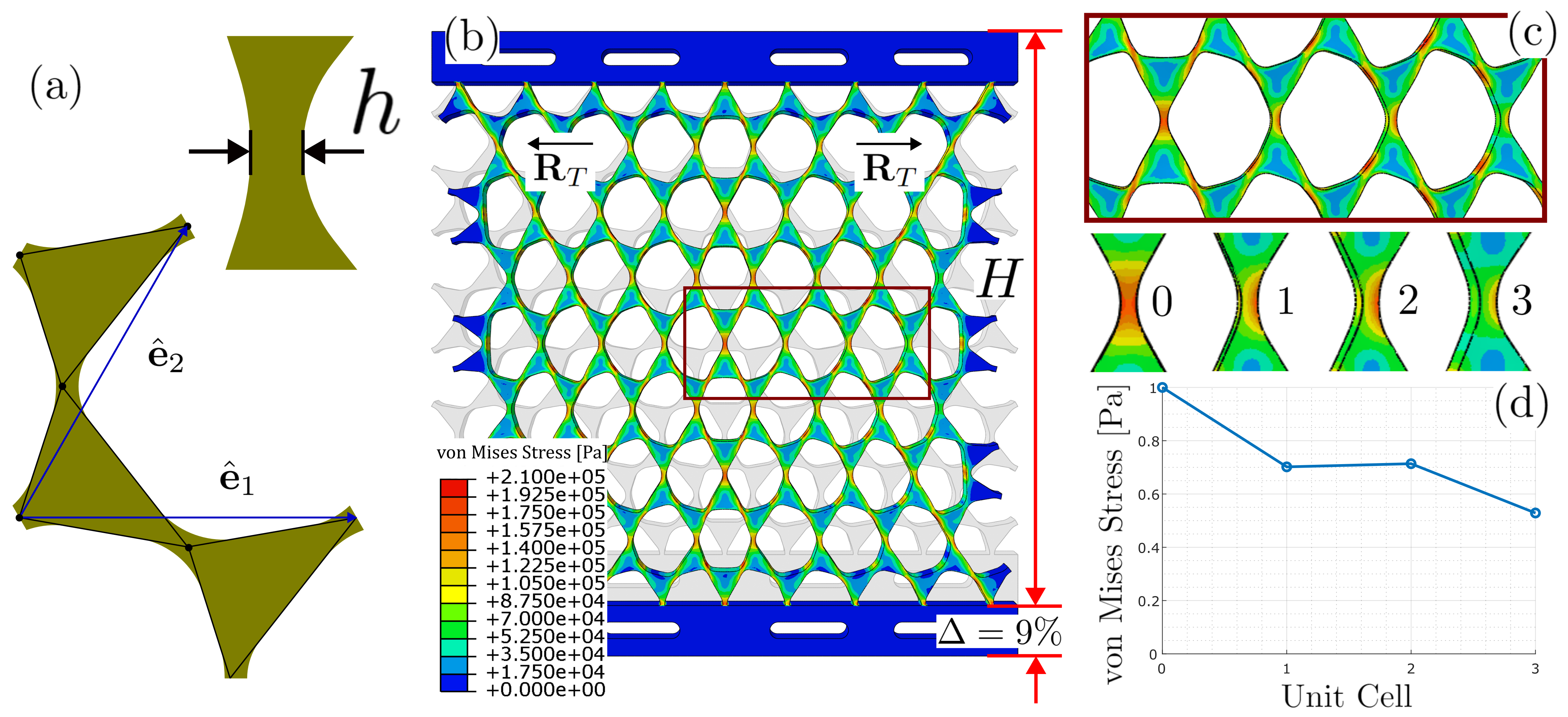}
	\caption{(a) Structural lattice unit cell with lattice vectors and finite thickness $h$ indicated. (b) Finite element (FE) model with von Mises stress field plotted for a $9\%$ total elongation. (c) Detail of the lattice  mid-line showing stress concentration at the domain wall. (d) Normalized stress versus cell index, demonstrating stress decay into the bulk to almost $50\%$ of the peak value.} \label{Soft_FEM}  
\end{figure*}

Recent works have addressed the resilience of the topological properties in the crossover from ideal configurations to structural lattices featuring finite-thickness hinges. Though structural lattices do not strictly meet Maxwell conditions due to the excess of constraints at the hinges, the signature of polarization is overall preserved, although the floppy modes migrate to finite frequencies and become soft phonons~\cite{JihongMa_2018,Stenull_2019,Charara_Bilayer_2021,Charara_PNAS_2022}, causing the polarization to manifest as asymmetric wave transport. In parallel, recent work by Kedia et al.~\cite{Kedia_2021} has proved the existence of soft static modes protected by topology and symmetry in hyperstatic lattices. 

Polarized lattices also exhibit the ability to localize states of self-stress (SSS)~\cite{KL_2013,Paulose_2015,Mao_Lubensky_2018,Chapuis_Shea_2022}. Kane and Lubensky first discussed the localization of SSS at interfaces, or \textit{domain walls}, between differently polarized sub-domains. Paulose et al. experimentally demonstrated the potential of SSS domain walls as stress guides to control the onset of selective buckling~\cite{Paulose_2015}. Zhang and Mao expanded the concept in the context of fracture protection, putting forth an implication of major engineering significance. When a conventional material with defects or cracks is loaded, the stress inevitably focuses in their neighborhoods, making them hot spots that can lead to failure. In contrast, when a lattice with an SSS domain wall is loaded, the stress focuses along the domain wall even in the presence of defects in the bulk and deep into the failure process, thus reducing the canonical stress concentration at the hot spots~\cite{Zhang_Fracture_2018}. 
Chapuis et al. also characterized topological Maxwell beam networks capable of localizing stress along non-linear domain wall interfaces~\cite{Chapuis_Shea_2022}. Finally, Widstrand et al.~\cite{Widstrand_2023} showed that stress focusing is preserved, albeit diluted in strength, in structural lattices with non-ideal hinges.  

The study of topological mechanics by and large has been framed within the bounds of linear elasticity, a few exceptions including the characterization of elastomeric polarized lattices in~\cite{Pishvar_Harne_2020} and recent work on soft lattices that use zero modes to achieve bulk shape reconfiguration~\cite{Jolly_2023,Hu_etal_2023}. 
Similarly, domain wall focusing has also been predominantly studied in the context of stiff lattices operating in the linear elastic regime. 
In this work, we revisit the problem in the realm of soft structures undergoing nonlinear finite deformation. 
Specifically, we address the following two practical and philosophical questions. 
1) Is stress focusing robust against the shape changes that the cells undergo during large deformation? What happens if we reach a deformation level that alters the topology of the bulk, forcing the lattice out of its polarized state, thus lifting the topological protection? 
2) Can we describe the dilution of polarization in terms of the evolution of some intuitive unit cell parameter? This would provide a powerful guideline to design structural lattices with desired stress management attributes forsaking the need for a precise characterization of the structural details. 

%\section{Soft metamaterial prototypes}

We conduct our investigation on a soft lattice prototype made of silicone rubber. This choice is dictated by practical considerations. Firstly, silicone features enough compliance to reach the deformation regimes required by our study without the onset of plasticity. Moreover, working with elastic moduli of the order of few MPa, the required deformation can be achieved under moderate tensile loads using dead weights, can be appreciated by naked-eye inspection, and can be quantified through nearly noise-free measurements via digital image correlation (DIC). This allows achieving a precise mechanical characterization working with a parsimonious table-top experimental set up. Secondly, since silicone has been widely used in mechanical metamaterials research, abundant literature is available, offering best practices for fabrication and characterization. 

\begin{figure*}[t] % !htb
	\centering
	\includegraphics[width=1\textwidth]{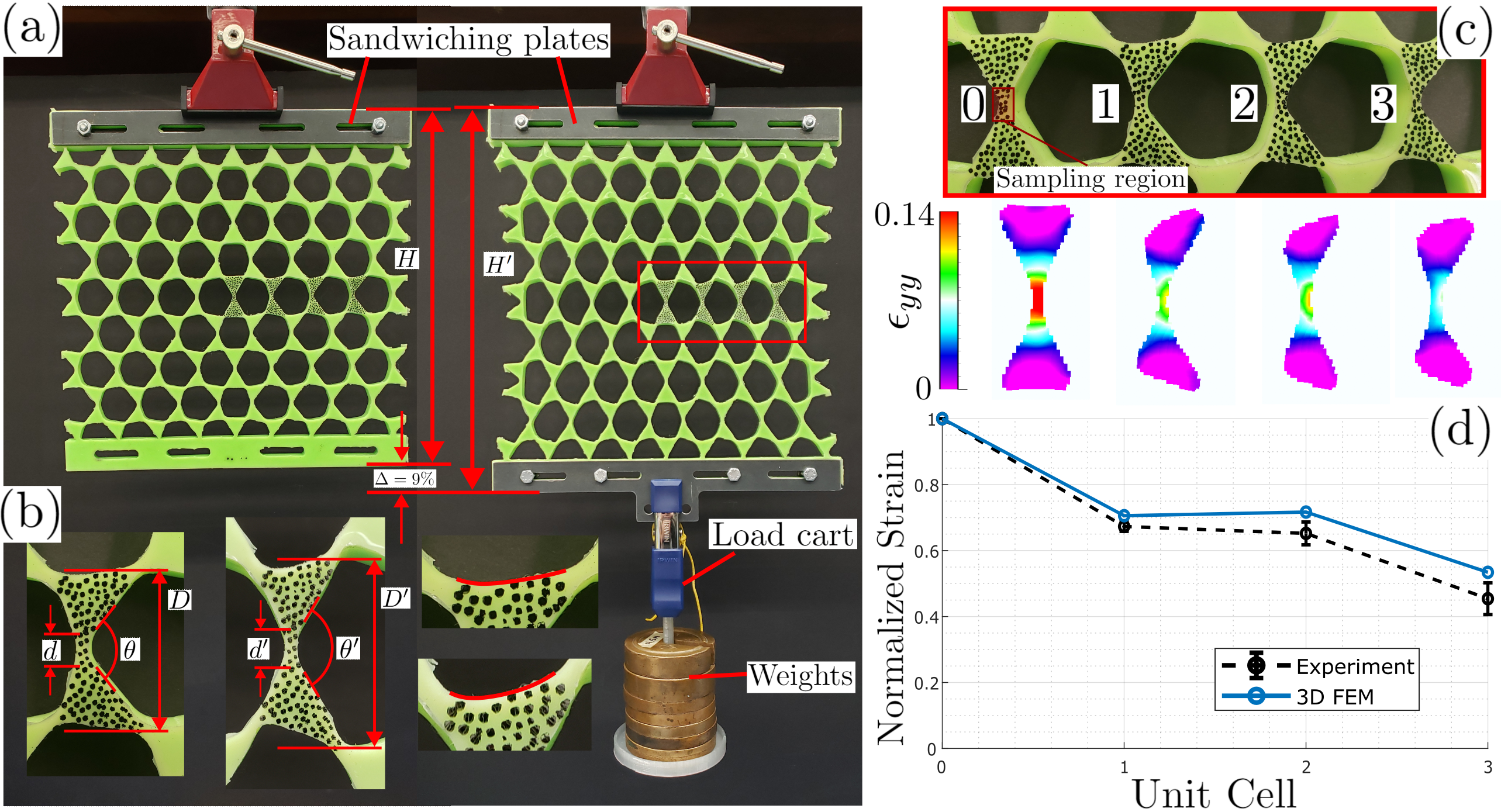}
	\caption{(a) Experimental set-up for uniaxial tensile testing of soft lattice specimens. Sandwiching plates grip and provide rigidity at the edges. 
    Dead load is applied at the bottom edge. (b) Evolution of the unit cell shapes under loading, involving changes in cell height, $D$, hinge ligament length $d$ and opening angle $\theta$, and onset of undulation of the edges, traced in red. (c) Detail of cells along the lattice mid-line, showing speckling pattern used for DIC and corresponding strain fields confirming strain concentrations at the domain wall (unit cell 0). (d) Plot of normalized strain versus cell index for experiments and FE, quantifying strain decay and focusing.} \label{Soft_Exp}  
\end{figure*}

Before discussing the tests conducted on the prototype, we develop a finite element (FE) model in Abaqus.
The selected configuration involves the distorted kagome cell shown in Fig.~\ref{Soft_FEM}(a). For ideal hinges, this configuration features a polarization vector pointing along primitive vector $\hat{\mathbf{e}}_1$. 
The domain, shown in Fig.~\ref{Soft_FEM}(b), consists of two polarized lattices, with their respective polarization vectors pointing outwards, stitched at their stiff edges to form an SSS domain wall. Note that the hinges are finite-thickness ligaments, making this a \textit{structural} kagome lattice. 
The FE model involves a mesh of eight-node brick elements, with geometric nonlinearity activated. To capture material nonlinearity, the silicone is modeled as a Yeoh hyperelastic material~\cite{Yeoh_1993,ABAQUS}. %with coefficients $C_{10} = 193884.95$, $C_{20} = -53015.04$, $C_{30} = 27143.08$ and $D_1 = 1.12\times 10^{-7}$. 
The coefficients are calibrated from stress-strain data obtained via uniaxial tension tests of dog-bone coupons made of the same material that we eventually use for our prototype (Zhermack Elite Double 32 silicone rubber, details in SM). 
The lattice is loaded uniaxially %tension by keeping the top edge fixed and
applying a tensile boundary condition to the bottom edge that results in an elongation of $\approx 9\%$, as shown in Fig.~\ref{Soft_FEM}(b).

From the von Mises stress field in Fig.~\ref{Soft_FEM}(b) we see that, while some stress concentration is observed in every hinge throughout the domain, the stress unequivocally localizes at the domain wall. In order to quantify the stress decay into the bulk, we extract stress values at the hinges (averaged over the hinge elements) at the domain wall as well as in the three immediately adjacent unit cells (Fig.~\ref{Soft_FEM}(c)) and plot them (normalized by the domain wall value) against the cell index in Fig.~\ref{Soft_FEM}(d). 
Within three cells, the stress decays to nearly half of its peak value, indicating substantial focusing even at this level of deformation. However, the strength of focusing is diluted when compared to data for a steel lattice operating in the linear regime under analogous tensile loading~\cite{Widstrand_2023}. 

\begin{figure*}[t] %!htb
	\centering
	\includegraphics[width=1\textwidth]{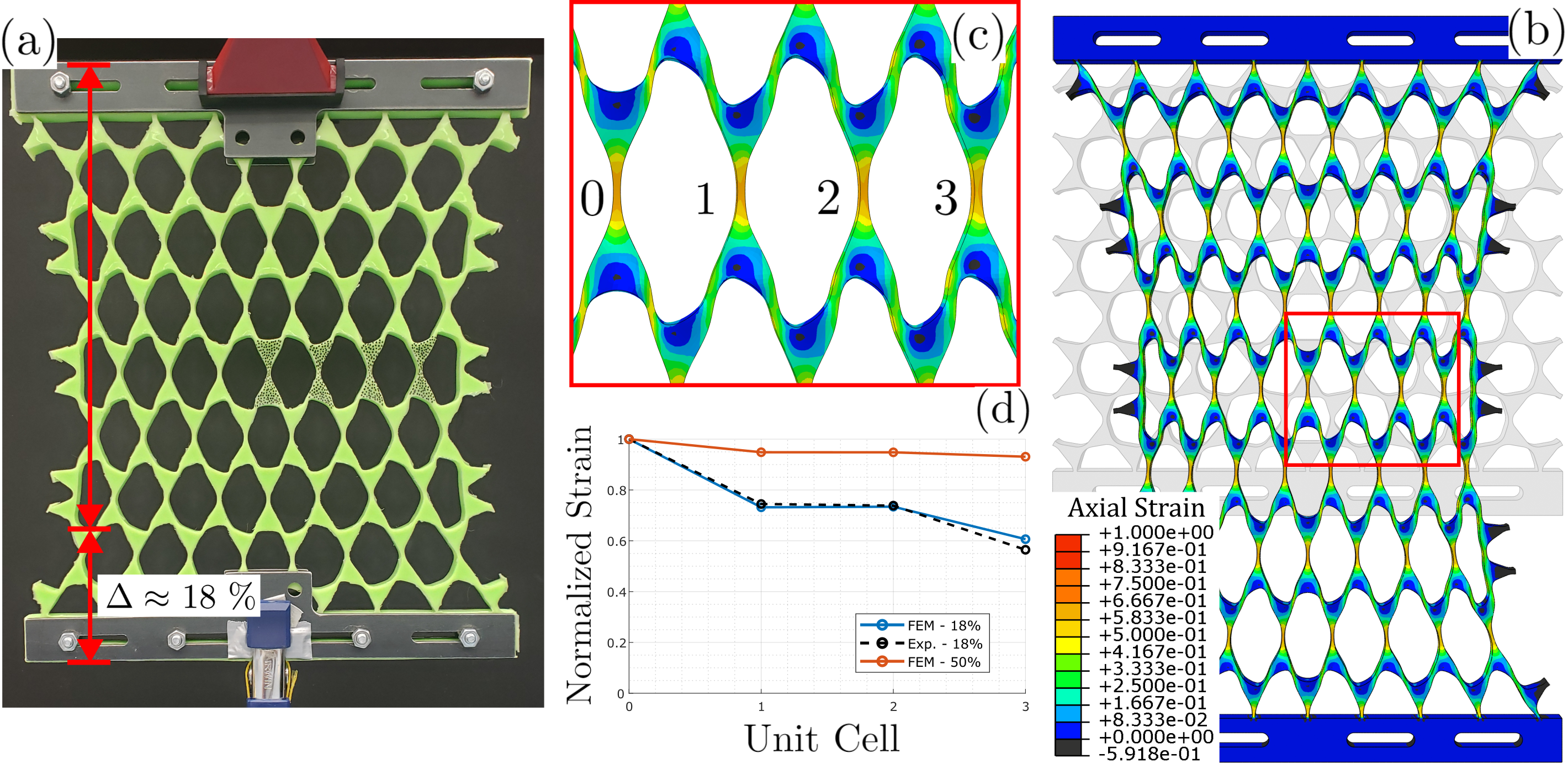}
	\caption{(a) Soft lattice specimen loaded to approximately 18\% total elongation. (b) FE analysis of soft lattice loaded to 50\% total elongation, with (c) detail of stress field showing loss of focusing. (d) Normalized axial strain vs. cell index for all cases.} \label{Extreme_Strain}  
\end{figure*}

We then proceed to seek experimental validation of these results. Following an established fabrication protocol, we fabricate our specimen by casting silicone in a 3D-printed mold, using the same silicone batch employed for the dog-bone coupons used for material characterization. We set the specimen in the dead-weight set-up shown in Fig.~\ref{Soft_Exp}(a). We sandwich the top and bottom edges between steel plates, which can be considered rigid, to ensure that the concentrated load is distributed uniformly along the loading edge, and we add weights until we reach $\approx$ 9\% of total elongation. 
At each loading stage, we take a snapshot of the speckle pattern in the region of interest, highlighted in Fig.~\ref{Soft_Exp}(c), and we feed the frames to a DIC software to infer displacements and axial strains (details in SM). The resulting strain map in Fig.~\ref{Soft_Exp}(c) reveals strong localization at the domain wall. To quantify the decay rate into the bulk, we plot the strains in the hinges, normalized by the peak value, against the cell index, superimposing the resulting curve to the one from the FE model in Fig.~\ref{Soft_Exp}(d). The results, displaying satisfactory agreement between experiments and simulations, confirm the rapid decay of the strain, which drops by over 50\% within three cells. From a visual inspection of the elongated specimen we infer a few morphological feature that will be valuable in later steps of our analysis. Specifically, it is clear from Fig.~\ref{Soft_Exp}(b) that the cells undergo a dramatic shape reconfiguration, where we can identify at least four mechanisms of deformation at work: 1) an elongation of the entire cell along the direction of loading, from $D$ to $D'$; 2) an axial stretching of the hinge from $d$ to $d'$; 3) a relative rotation of the triangles from $\theta$ to $\theta'$; and a loss of straightness of the edges. 

To test whether the effect remains appreciable under more extreme deformation, we double the elongation to 18\%, producing the deformation shown in Fig.~\ref{Extreme_Strain}(a) accompanied by a more pronounced degree of shape reconfiguration of the cells. The plot of axial strain vs. unit cell in Fig.~\ref{Extreme_Strain}(d) 
shows that a considerable amount of focusing is maintained even under this level of elongation, with strain dropping by 40\% within three cells. In order to force a loss of polarization, 
we further increase the load to reach a total elongation of 50\%. Since our lab setup can only handle elongations up to 30\%, for this case we resort to FE calculations, whose reliability was verified in previous loading cases. From the strain field in Fig.~\ref{Extreme_Strain}(b-c) and the orange curve in Fig.~\ref{Extreme_Strain}(d), we can see that the focusing is now completely lost, with the normalized value approaching 1 far from the domain wall. 

%\section{Cell reconfiguration and the sources of polarization dilution}

\begin{figure*}[t]
	\centering
	\includegraphics[width=1\textwidth]{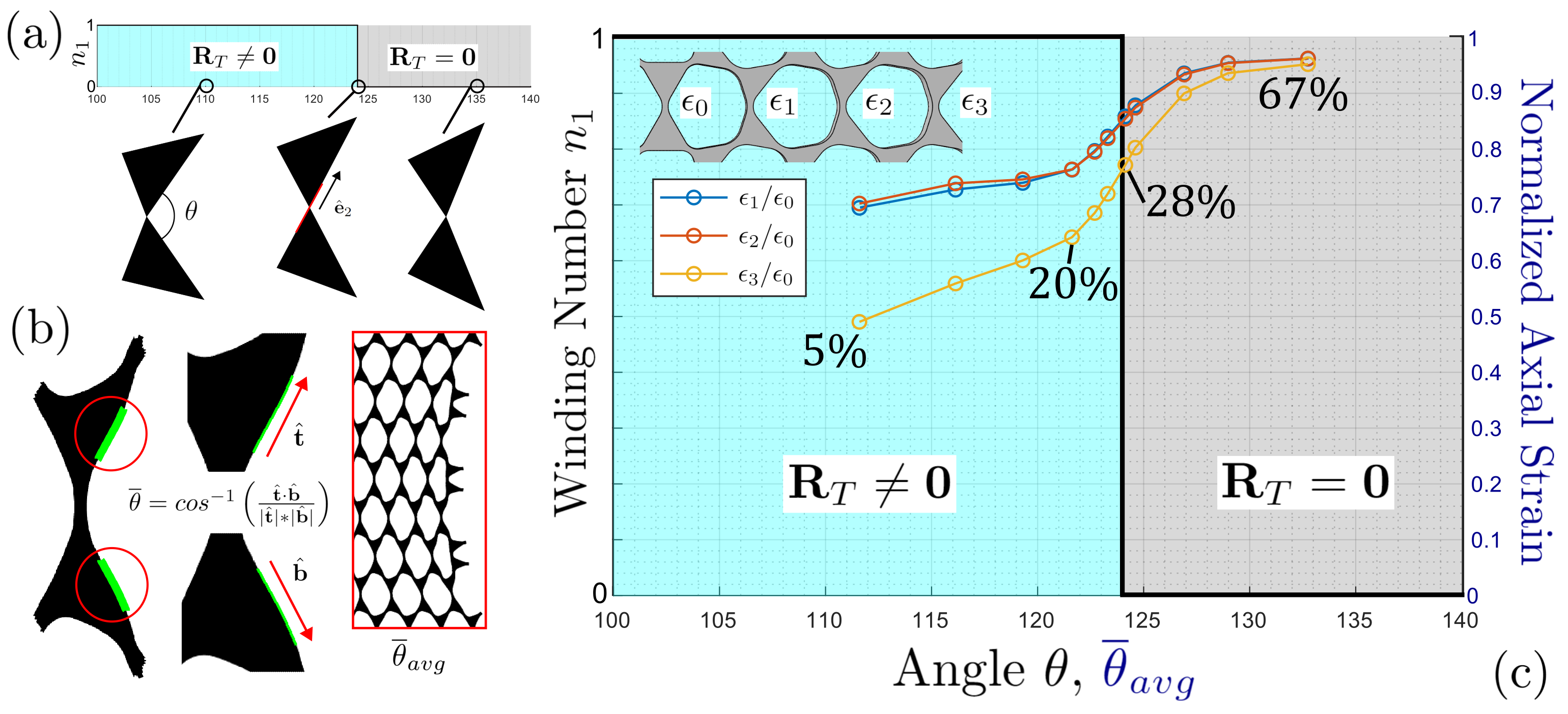}
	\caption{(a) Winding number diagram for the idealized unit cell. (b) Structural unit cell with traced edges and effective angle $\overline{\theta}$ computed from the vectorized traced edges, averaging over 25 cells to filter out spatial variability. \sg{(c)} FE-enabled sweep of load and strain levels in hinges at, close and further away from the domain wall, for the structural lattice. The evolution is superimposed on the winding number diagram for the ideal case, letting the average effective opening angle $\overline{\theta}_{avg}$ run along the same axis of the ideal angle $\theta$. The curves show a progressive decay of focusing with increasing load levels and $\overline{\theta}_{avg}$, with total loss for elongations above 50\%. The focusing drops rapidly over a relatively narrow window of stages centered around 28\% elongation, which correlates with the critical value of $\theta$ marking the topological phase transition for the ideal lattice.} \label{Winding_Strain}  
\end{figure*}

%Our next step is to find a connection between the loss of polarization observed in soft structural lattices and the evolution of one or more key parameters of the unit cell that are swept during loading. 
%We recall that the cell deformation involves at least four major kinematic contributions: total cell elongation, stretching of the hinges, rotations of the triangles and loss of straightness of the edges. While it is difficult to precisely rank these effects in terms of their contribution to the dilution of polarization, we choose to focus our attention on the rotation $\theta$ which, among these factors, is the only one that also plays a role in the kinematics of ideal lattices. In fact, $\theta$ fully captures the Guest-Hutchinson mode~\cite{Guest_2003}, which reconfigures ideal Maxwell lattices without costing elastic energy~\cite{Sun_Kagome_PNAS_2022}. Therefore, a parameterization in terms of $\theta$ allows a precious description of the evolution of polarization in terms of an \textit{interpretable} parameter. 

\sg{Our next step is to find a connection between the loss of polarization observed in soft structural lattices and the evolution of one or more key parameters of the unit cell that are swept during loading. We recall that the cell deformation involves at least four major kinematic contributions: total cell elongation, stretching of the hinges, rotations of the triangles and loss of straightness of the edges. While it is difficult to precisely rank these effects in terms of their contribution to the dilution of polarization, we choose to focus our attention on the rotation $\theta$ which, among these factors, is the only one that also plays a role in the kinematics of ideal lattices. Therefore, a parameterization in terms of $\theta$ is chosen deliberately because it allows a precious description of the evolution of polarization in terms of an \textit{interpretable} parameter with an explicit and formal connection to the topology of the corresponding ideal lattice.} 

%Incidentally, a $\theta$ sweep fully captures the kagome lattice Guest-Hutchinson mode~\cite{Guest_2003}, i.e., the nonlinear global soft mode that reconfigures an ideal kagome lattice~\cite{Sun_Kagome_PNAS_2022}.

Here we build upon the work by Rocklin et al.~\cite{Rocklin_2017}, in which it was shown that, for a given ideal kagome geometry, it is possible to turn on and off the polarization through a uniform twist described by an angle $\theta$. Note that, from a design perspective, varying $\theta$ can be interpreted as a sweep of the single-variable design space of the given kagome family. However, it can also be interpreted as a step-by-step illustration of the active reconfiguration sweep that the lattice naturally undergoes while experiencing a global soft mode, known as a Guest-Hutchinson mode~\cite{Guest_2003}. Such reconfiguration brings about two topological phase transitions between unpolarized to polarized states, %. Specifically, as $\theta$ increases, the lattice first undergoes a transition from unpolarized to polarized, and then a second one back to the unpolarized state. The transitions 
which are marked by special configurations, in which the bonds are aligned, giving rise to continuous ``fibers" that support states of self stress (SSS). 
Recently, similar notions have been applied to multi-stable lattices, where external loading causes a change in the morphology of the bulk that results in a topological phase transition~\cite{Xiu_2022}.

We leverage this framework to propose a connection between the dilution of stress focusing in the structural lattice and the topological phase transition in its ideal counterpart. For an ideal lattice, we can determine the polarization by computing two topologically invariant winding numbers $n_1$ and $n_2$ which capture the decay, or lack thereof, of wave modes along  
the directions of $\hat{\textbf{e}}_1$ and $\hat{\textbf{e}}_2$, respectively. A linear combination of $\hat{\textbf{e}}_1$ and $\hat{\textbf{e}}_2$, with $n_1$ and $n_2$ serving as coefficients, yields the the polarization vector $\textbf{R}_T$, which captures the polarization encoded in the bulk topology~\cite{KL_2013,Chapuis_Shea_2022}. For the geometry in Fig.~\ref{Winding_Strain}(a), parameterized in terms of $\theta$, $n_2=0$ identically, while $n_1$ undergoes an abrupt jump from 1 to 0 for $\theta \approx 124^o$, which marks a topological phase transition from polarized ($\textbf{R}_T \neq 0$) to unpolarized ($\textbf{R}_T=0$), denoted by the cyan- and gray-shaded areas. Interestingly, this switch corresponds to an alignment of the edges of the triangles along $\hat{\textbf{e}}_2$, as predicted in~\cite{Rocklin_2017}.

Before we can perform a meaningful comparison between the value of $\theta$ at which topological phase transition occurs in the ideal lattice and the effective rotation conditions for which the dilution of polarization manifests in the structural lattice, we must address some kinematic ambiguities embedded in the large deformation field. First, since the edges of highly deformed triangles are undulated, it is challenging to precisely define an opening angle $\theta$, and we must instead resort to an \textit{effective opening angle} $\overline{\theta}$. To this end, we trace the boundaries of the deformed cell using an image processing software (details in SM), we approximate the traced edges locally (close to the hinges) with segments described by vectors $\hat{\textbf{t}}$ and $\hat{\textbf{b}}$, and we compute $\overline{\theta}$ from their dot product, as shown in Fig.~\ref{Winding_Strain}(b). Furthermore, we note that the effective opening angle is highly variable across the domain, due to the severe non-uniformity of the deformation field. To account for this variability, we average the values of $\overline{\theta}$ inferred from a sample window of cells (red box in Fig.~\ref{Winding_Strain}(b)) to extract an \textit{average} effective opening angle $\overline{\theta}_{avg}$.

 \begin{figure*}[t] %!htb
	\centering
	\includegraphics[width=1\textwidth]{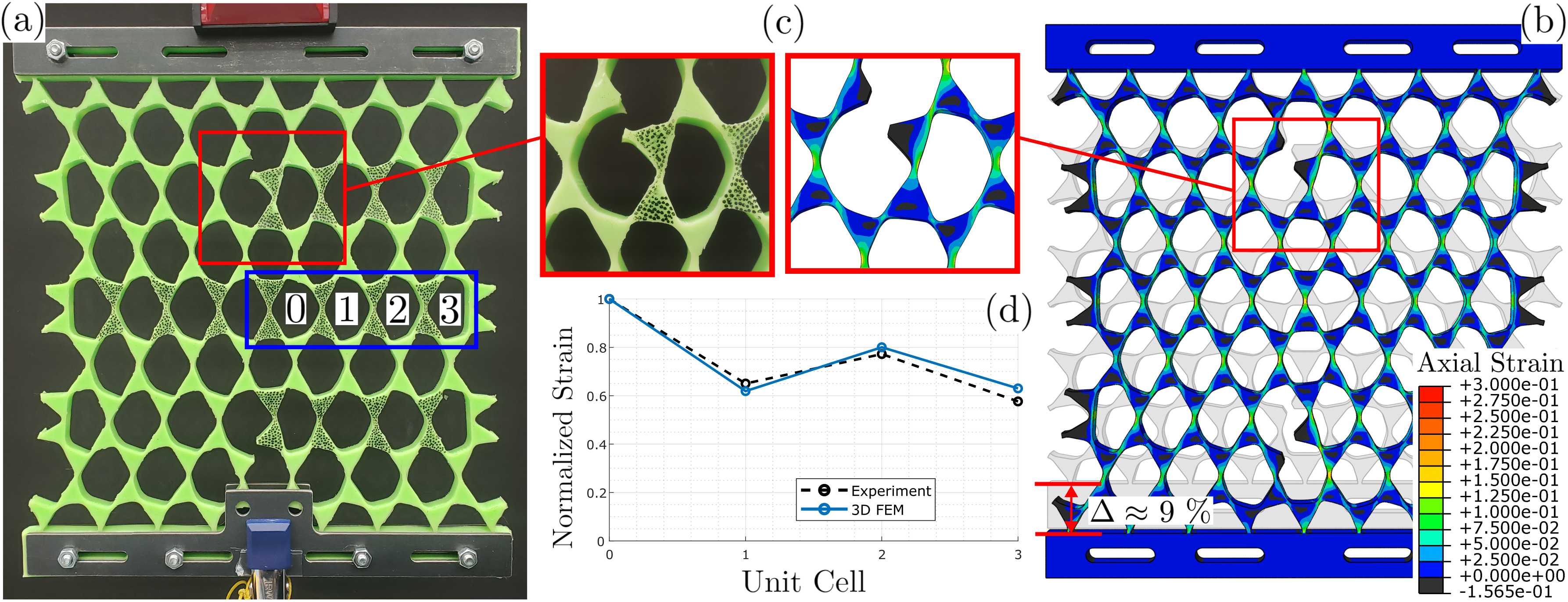}
	\caption{(a) Experimental response of the soft lattice specimen with two cuts along the domain wall. (b) Finite element model of the damaged lattice showing overall preservation of the focusing effect in the axial strain field. (c) Details of the experimental and simulated deformation fields showing the rearrangement of deformation in the neighborhood of the defected hinges that stop carrying loads. (d) Normalized axial strain vs. cell index for experiments and FE. The curve shows some local pick-up in strain in unit cell 2, but overall focusing at the domain wall is maintained.} \label{TwoCutsResult}  
\end{figure*}

In our FE model, we simulate increasing levels of elongation (from 5\% to 67\%) which result in increasing values of $\overline{\theta}_{avg}$. For each level, we compute the hinge axial strain in three cells located one, two and three cell positions from the domain wall, respectively, and we normalize the values by that at the domain wall. In Fig.~\ref{Winding_Strain}(b) we plot the normalized strains vs. $\overline{\theta}_{avg}$, superimposing the curves to the winding number diagram for the ideal case, deliberately sweeping $\overline{\theta}_{avg}$ and $\theta$ on the same axis in order to put the two quantities in direct comparison. We observe a rise of the normalized strain away from the domain wall, tending to 1 for extreme loads, confirming a progressive dilution of the polarization as the load increases. Interestingly, inspecting the curves against the backdrop of the winding number, most of the increase occurs in a relatively narrow range of $\overline{\theta}_{avg}$ values clustered around an inflection point that matches the critical value of $\theta$ that marks the phase transition for the ideal lattice. This suggests that a simple kinematic descriptor of the ideal lattice can serve as a viable predictor of stress focusing even away from ideal lattice conditions. This observation has two profound implications. On one hand, it unequivocally links the change in focusing of the structural lattice to the topological character of its ideal counterpart. On the other hand, it provides a powerful design guideline, in that we can predict some key features of a structural configuration regardless of the availability of structural details about the hinges, only relying on information about geometry and connectivity. 

%\section{Robustness of focusing against onset of defects along the Domain Wall}

Our last goal is to verify the robustness of focusing against the onset of damage. In previous work, we had already demonstrated significant robustness in going from ideal to structural lattices, albeit limited to operating in the linear elastic regime~\cite{Widstrand_2023}. We now assess whether such protection persists under large deformation. Again, we conduct the assessment experimentally, see Fig.~\ref{TwoCutsResult}(a), and via simulations, see Fig.~\ref{TwoCutsResult}(b). We introduce damage by severing two hinge ligaments along the domain wall (insets in Fig.~\ref{TwoCutsResult}(c)) and we determine the resulting strain field to assess the existence of decay patterns. We observe that the strains in hinges labeled 0-3, moving away from the domain wall, retain a substantial decay. From Fig.~\ref{TwoCutsResult}(b), we can appreciate how the largest strains remain confined along a wavy path that bounds the defected domain wall, always peaking at the first available undamaged hinge. The normalized strain plotted in Fig.~\ref{TwoCutsResult}(d) quantifies the persistence of the localization, whereby, 
while the defect appears to cause a pickup in strain in the second cell, the local increase is not enough to overcome the overall focusing ability afforded by the domain wall. 

%\section{Conclusions}
In conclusion, we demonstrated experimentally the availability of stress-focusing in soft structural kagome lattices experiencing finite deformation. The focusing remains substantial even under elongation levels that produce significant morphological changes in the cells. Furthermore we determined that the twist angle for ideal lattices remains a good predictor of focusing for structural lattices with more complex hinge mechanics.

%\section{Acknowledgments}
This work is supported by the National Science Foundation (awards CMMI-2027000 and CMMI-2026794), The authors acknowledge support from the Minnesota Supercomputing Institute and the UMN Anderson Labs and are grateful to J. Labuz (UMN) for help with DIC equipment and expertise. 

\bibliographystyle{apsrev4-2}
\bibliography{Kagome_Bib} 

\onecolumngrid

\newpage

\setcounter{figure}{0}

\section{SUPPLEMENTAL MATERIAL: Robustness of stress focusing in soft lattices under topology-switching deformation}

\subsection{Uniaxial testing and coefficients of hyperelastic material model}
\begin{figure*}[h]
	\centering
	\includegraphics[width=0.8\textwidth]{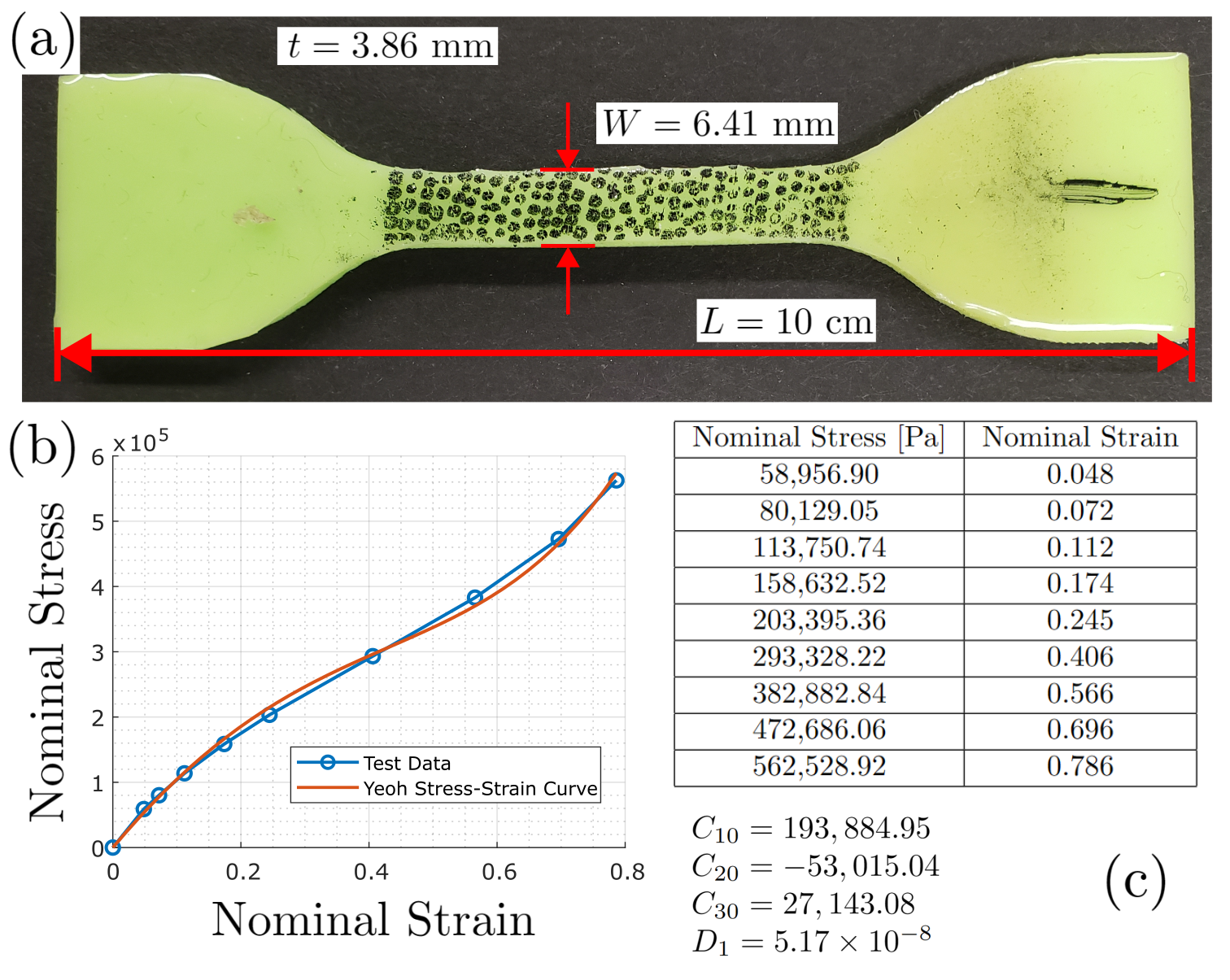}
	\caption{(a) Silicone rubber unixial testing coupon. A speckle pattern is drawn on the surface of the lattice to compute strain via DIC. (b) Nominal strain vs nominal stress plot. Test data is averaged across several tests of the coupon. The Yeoh stress-stran curve is obtained by fitting the data to the Yeoh mdoel in Abaqus. (c) Tabulated stress-strain data and corresponding Yeoh model coefficients.} \label{Uniaxial_Tests}  
\end{figure*}

The hyperelastic material model used in the finite element simulations is determined from uniaxial test data conducted on a silicone rubber coupon cast from the same batch of material used to cast the lattice specimen. 
The silicone rubber coupon is pictured in Fig.~\ref{Uniaxial_Tests}(a). 
The specimen is loaded with increasingly larger dead loads, similar to the testing procedure used on the full lattice specimen. 
These loads are recorded as nominal stress in the table pictured in Fig.~\ref{Uniaxial_Tests}(c). 
Nominal strain is computed using digital image correlation (DIC) on the speckle pattern adorning the surface of the coupon in Fig.~\ref{Uniaxial_Tests}(a). 
Nominal strain is tabulated alongside nominal stress and this stress-strain curve is also plotted in Fig.~\ref{Uniaxial_Tests}(b). 
Using the nominal stress-strain curve as input, and assuming we have a nearly incompressible material ($\nu = 0.495$), we use the finite element software Abaqus to fit the test data to a Yeoh hyperelastic material model. 
The hyperelastic stress-strain relationship is plotted over the range of test data in Fig.~\ref{Uniaxial_Tests}(b) and we can see that the Yeoh model closely fits the test data for this level of strain. 
The coefficients of the Yeoh material model are given in Fig.~\ref{Uniaxial_Tests}(b). 
To run the finite element analysis discussed in the main text, these coefficients are used. 

\newpage
\subsection{Image-Processing Procedure for Unit Cell Boundaries}
In order to trace the boundaries of the lattice's unit cells, we have written and implemented an image-processing code. We define a subset of unit cells in the bulk of the lattice over which we iterate the code. For each iteration, we plot an individual deformed unit cell and binarize the image so that it can be stored as a binary matrix where the zeroes correspond to pixels outside of the boundaries of the unit cell and the ones correspond pixels within the boundaries of the unit cell. Since we must locate the \textit{upper} and \textit{lower} triangles of the unit cell, we search through two rows of the matrix from right-to-left for the first value of \textit{one} in the row. The pixel corresponding to that one becomes the seeding point for the MATLAB function \textit{bwtraceboundary}. This function enables us to trace the outline of a binarized image, and it also provides us with the coordinates of the points along the traced path. We do this in two locations in order to trace a short path along both the upper and lower triangles of the unit cell. Using the coordinates of these paths, we fit vectors to the paths and compute the angle between these vectors using their dot product.

\end{document}